\documentclass{moriond}

\def\be{\begin{equation}}
\def\ee{\end{equation}}
\def\bea{\begin{eqnarray}}
\def\eea{\end{eqnarray}}

\usepackage{subcaption}

\newcommand{\rl}{$R_{\rm L}$}
\newcommand{\ptrl}{$\langle p_{\rm T} \rangle R_{\rm L}$}
\newcommand{\pt}{$p_{\rm T}$}
\newcommand{\ptavg}{$\langle p_{\rm T} \rangle$}
\newcommand{\gev}{GeV/\textit{c}}
\newcommand{\Dzero}{$\rm{D}^{0}$}
\newcommand{\kt}{$k_{\rm T}$}

\newcommand{\Photo}{\includegraphics[height=35mm]{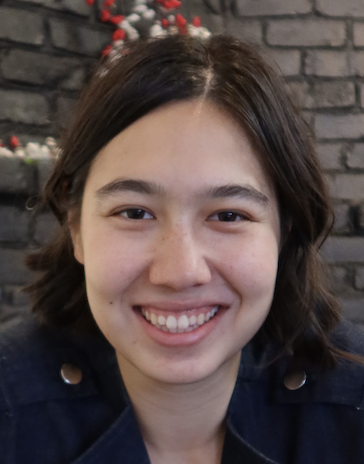}}

\begin{document}
\vspace*{4cm}
\title{Energy-energy correlators in small and large systems}

\author{ Beatrice Liang-Gilman on behalf of the ALICE Collaboration}

\address{Department of Physics, University of California Berkeley,\\
Berkeley, California, 94709, USA}

\maketitle\abstracts{
Energy-energy correlators (EECs) provide a powerful tool to study the evolution of scattered partons into final-state hadrons. In these proceedings, a variety of energy correlator measurements performed by the ALICE collaboration are reported. The 2-point energy-energy correlator (EEC) is measured in inclusive jets and heavy-flavor jets in pp collisions, as well as in inclusive jets in p--Pb collisions. The 3-point energy correlator is also discussed, including prospects for its use in extracting the strong coupling constant.
}

\section{Introduction to energy correlators}

Jets are a powerful probe for studying Quantum Chromodynamics (QCD) across various systems. The energy correlator (ENC) observable is a jet substructure tool that characterizes energy flow within a jet. Presented here are a few of the ENC measurements performed by the ALICE experiment, covering various types of jets and collision systems.

The $N$-point energy correlator is defined as the energy-weighted cross section of the angle between $N$ particles.

\be
\Sigma_{\rm ENC}(R_{\rm L}) = \frac{1}{N_{\rm jet} \cdot \Delta}\int_{R_{\rm L} - \frac12 \Delta}^{R_{\rm L} + \frac12 \Delta} \sum_{\rm jets} \sum_{i,j} \frac{p_{{\rm T}, i} p_{{\rm T}, j}...p_{{\rm T}, N}}{p_{\rm T, jet}^N}\delta(R_{\rm L}' - R_{{\rm L}, ij}){\rm d}R_{\rm L}',
\label{eq:eec}
\ee
where $N$ is the number of particles used to construct the correlator, \rl~is the angular distance between any two particles, defined as $\sqrt{\Delta\varphi_{ij}^2 + \Delta\eta_{ij}^2}$, and the energy weight is defined as $p_{{\rm T},i}p_{{\rm T},j}...p_{{\rm T},N}/p_{{\rm T, jet}}^N$. The ENC shape shows a clear separation between the perturbative and non-perturbative regimes, with a transition region at the peak that is sensitive to \mbox{ hadronization. \cite{eec_intro_chen}}

\section{2-point energy-energy correlators (EEC)}

\subsection{EEC in pp collisions}

\begin{figure}[h!]
    \begin{subfigure}{0.49\textwidth}
        \centering
        \includegraphics[scale=.31]{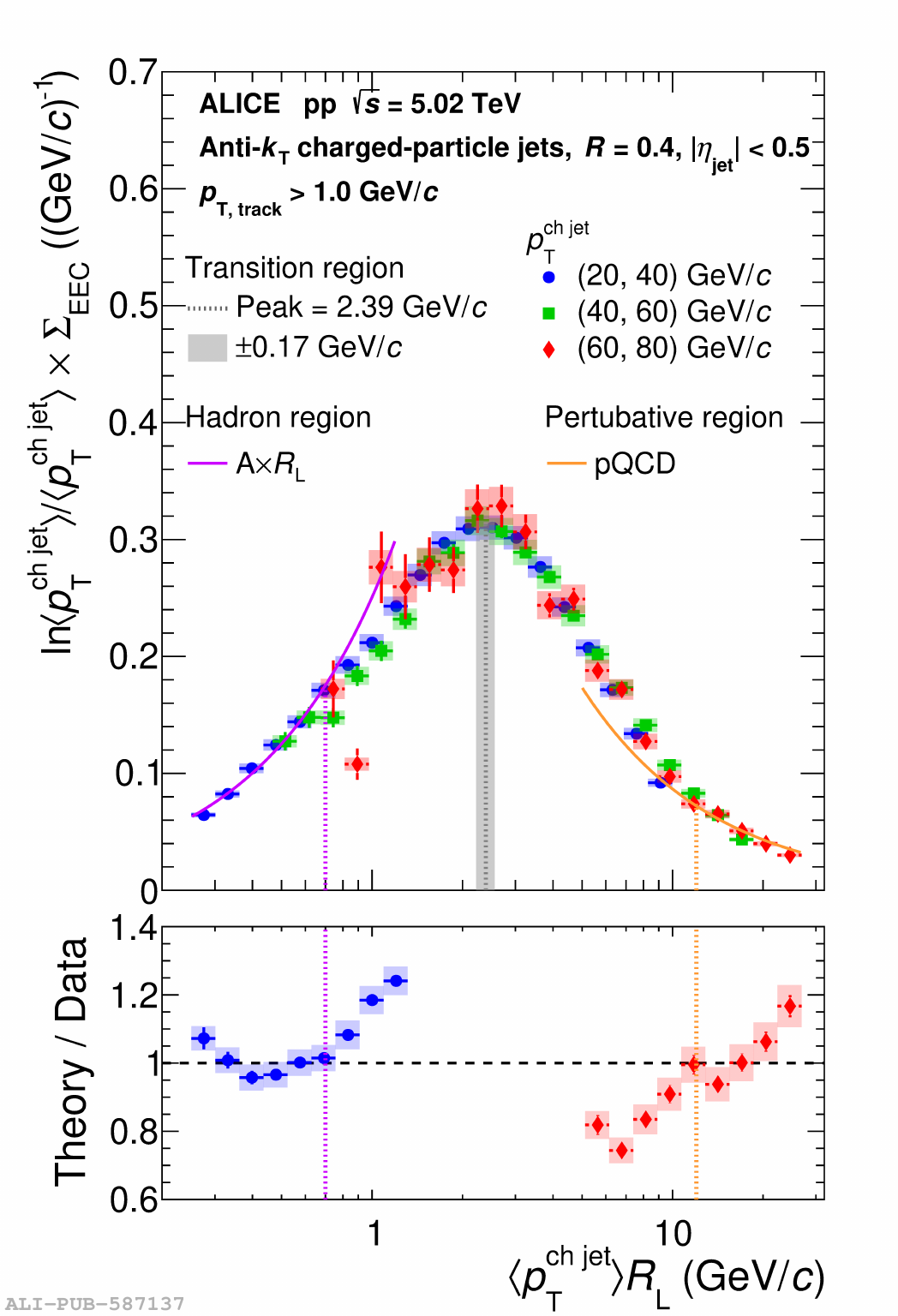}
    \end{subfigure}
    \begin{subfigure}{0.49\textwidth}
        \centering
        \includegraphics[scale=.6,trim={0 0 10cm 0},clip]{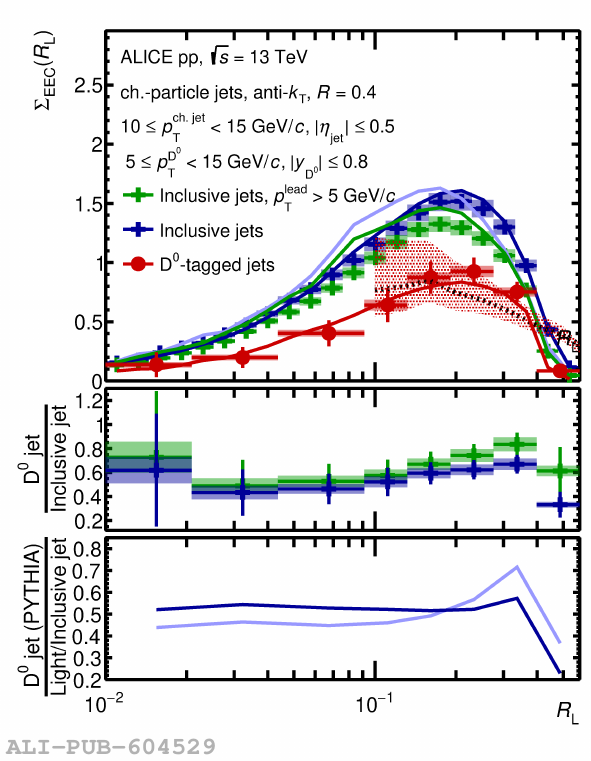}
    \end{subfigure}
    \caption{\textbf{(a)} The EEC in pp collisions, plotted as a function of \ptrl~for three different jet \pt~ranges. \textbf{(b)} The $D^0$-jet EEC compared to inclusive jet EEC shown for 10-15 \gev~jets.} 
    \label{fig:eec_pp}
\end{figure}

The simplest form of the $N$-point energy correlator is the 2-point correlator, otherwise known as the energy-energy correlator (EEC). The inclusive jet energy-energy correlator measured by ALICE in pp collisions for $R=0.4$ anti-\kt~charged jets is shown in Fig.~\ref{fig:eec_pp}(a) for three different jet transverse momentum ranges. The EEC is plotted as a function of \ptrl, where \ptavg~is the average jet \pt~per \pt~range studied. The right side of the plot features wider angle pairs and corresponds to the perturbative region of the EEC, associated with early time splittings. In contrast, the left side of the figure contains smaller angle pairs and corresponds to the non-perturbative region, where free hadron scaling becomes relevant. The peak in the middle corresponds to the transition region. When the EEC is plotted as a function of \ptrl, it exhibits alignment across different jet momenta, revealing a jet \pt-independent universality in jet dynamics. \cite{alice_eec_pp} 

\subsection{Heavy-flavor EEC in pp collisions}

Charm-tagged jets were also studied using the EEC in order to characterize the radiation pattern of charm-initiated showers. Heavy-flavor quarks are an effective probe for studying QCD as they are created in the initial stages of high-energy collisions. When a quark of mass $m$ and energy $E$ propagates after the initial scattering, the pattern of the parton shower is expected to depend on the quark mass through a phenomenon known as the dead-cone effect, where radiation from the quark is suppressed at angular scales smaller than $m/E$ around the direction of the quark. \cite{alice_deadcone}

ALICE studied charm-tagged jets via a \Dzero~hadron using the experiment's excellent tracking resolution and particle identification capabilities. The \Dzero~meson, which contains a charm quark, was reconstructed from its $K^{\mp}\pi^{\pm}$ decay channel. The \Dzero-tagged jets were obtained by first reconstructing the \Dzero~candidates, clustering the charged particles with the reconstructed neutral \Dzero~mesons into anti-\kt~jets of $R=0.4$, and finally selecting the jets that contained a \Dzero~meson.

The \Dzero-jet EEC is shown in Fig.~\ref{fig:eec_pp}(b). The \Dzero-tagged jets, in the red points, are visibly suppressed compared to the inclusive jets, in the dark blue points. The reduced yield corresponds to a reduction of the number of pairs, which is consistent with the expectations of suppressed radiation due to the dead cone. \cite{alice_eec_djets} In addition, the peak position of the inclusive jet EEC is consistent within 1$\sigma$ of the peak position of the \Dzero-tagged jet EEC. This is due to the interplay of mass and color effects. The flavor effects arise from the Casimir color factors of the quarks and gluons, which shift the peak position for inclusive jets to larger \rl~relative to light-quark jets, while the quark-mass effects appear as the dead cone. There is also some indication of the importance of non-perturbative effects such as hadronization, given the slight tension between the scaling of the pQCD calculation \cite{eec_croft_pythia} and the data.

\subsection{EEC in p--Pb collisions}

Energy-energy correlators can also provide insights into how jets differ across various collision systems. ALICE studied inclusive jets in p--Pb collisions to investigate possible jet modification due to cold nuclear matter effects. In this environment, initial state effects, such as the presence of the nPDF, and final state effects, such as multiple scattering or comovers, could be present.

The background in p--Pb collisions is more significant than in pp collisions, so an underlying event (UE) subtraction is performed. This analysis quantified the UE contribution to the jet energy using an area-corrected median subtraction method. \cite{cms_rho_sub} Then, the UE-related combinatorial background in the EEC is corrected for using a perpendicular-cone method, which uses a cone 90$^\circ$ in $\varphi$ away from the signal jet to estimate the background contribution. 

\begin{figure}[h!]
    \begin{subfigure}{0.49\textwidth}
        \centering
        \includegraphics[scale=.35]{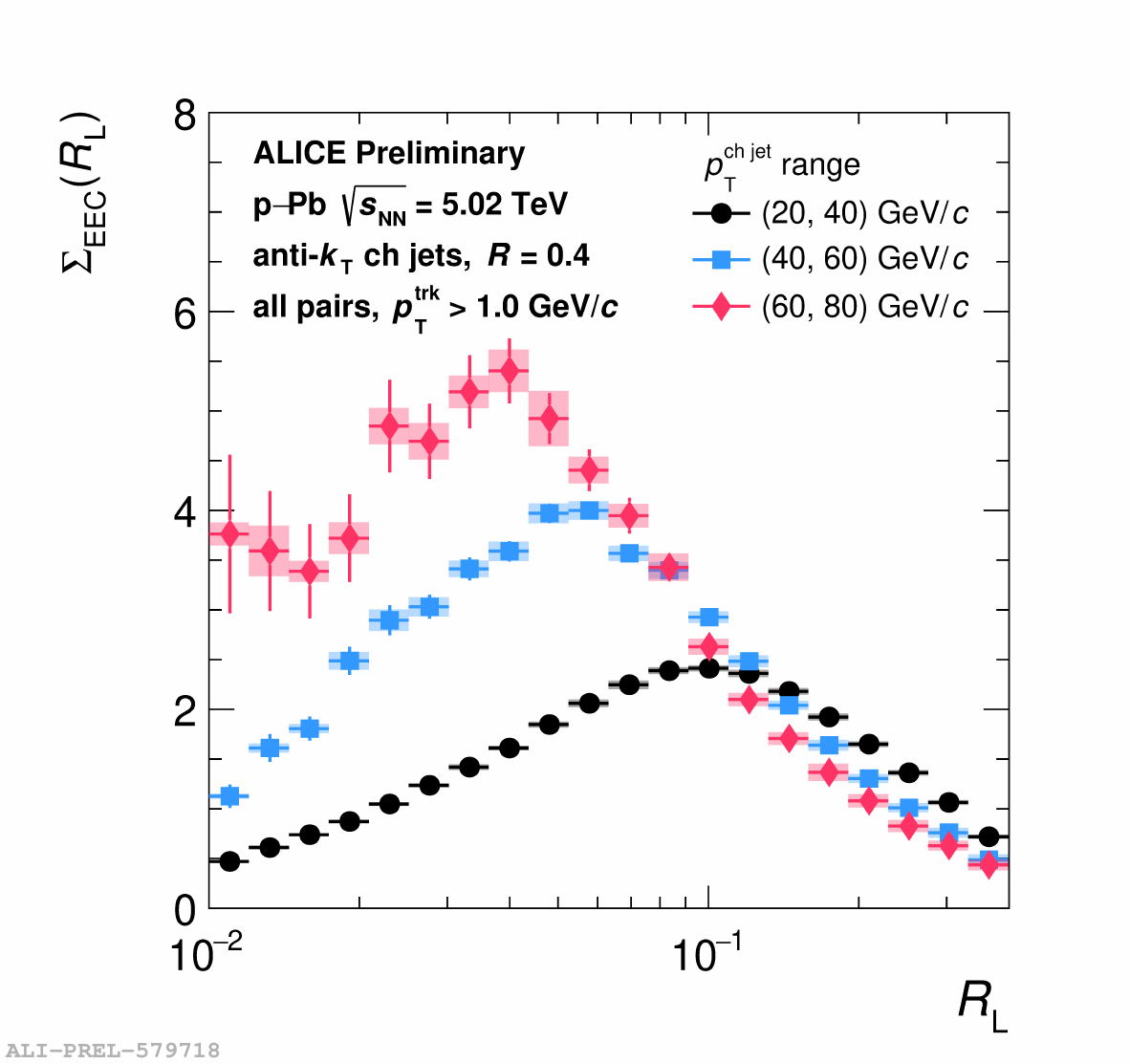}
    \end{subfigure}
    \begin{subfigure}{0.49\textwidth}
        \centering
        \includegraphics[scale=.35]{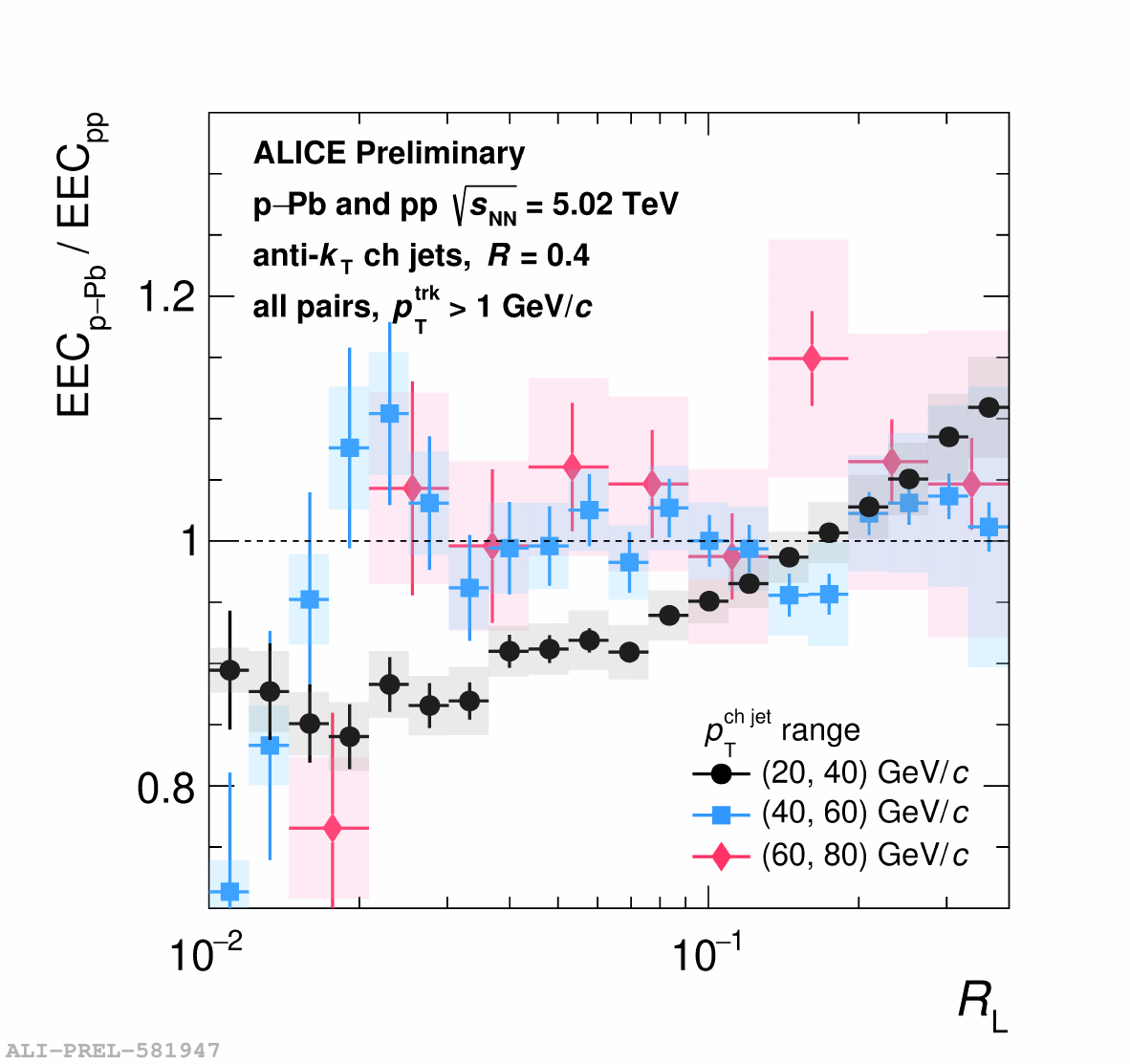}
    \end{subfigure}
    \caption{\textbf{(a)} The EEC of jets in p--Pb collisions for three different jet momenta and \textbf{(b)} the ratio of the EEC in p--Pb collisions to the EEC in pp collisions in the same jet \pt~ranges.}
    \label{fig:eec_ppb}
\end{figure}

Figure~\ref{fig:eec_ppb}(a) shows the EEC of inclusive charged-particle anti-\kt~jets with $R=0.4$ in \mbox{p--Pb} collisions for three different jet \pt~ranges. To identify any differences between p--Pb and pp jets, the ratio of the corresponding EECs is taken, as shown in Fig.~\ref{fig:eec_ppb}(b). The only jets that show modification are those in the lowest jet \pt~range, 20--40 \gev. These jets reveal an approximately 10\% suppression in the small-angle region, and an approximately 10\% enhancement in the large-angle region. While some theory calculations were performed to try to explain these differences \cite{ppb_theory_fu} \cite{ppb_theory_barata} \cite{ppb_theory_andres}, the underlying cause of this modification remains currently undetermined. Recent work points to the potential of attributing this modification to an increase in jet constituent multiplicity for jets in p--Pb compared to pp collisions, however this work is still underway. \cite{anjali_QM25_slides}

\section{3-point energy correlators}

The 3-point energy correlator, or E3C, probes higher order energy flow correlations, and can also be used to extract the strong coupling constant. \cite{e3c_theory} \cite{cms_e3c} The E3C is calculated by finding the largest angular distance, \rl, between all triplets of tracks in a jet.

\begin{figure}[h!]
    \begin{subfigure}{0.49\textwidth}
    \centering
        \includegraphics[scale=.35]{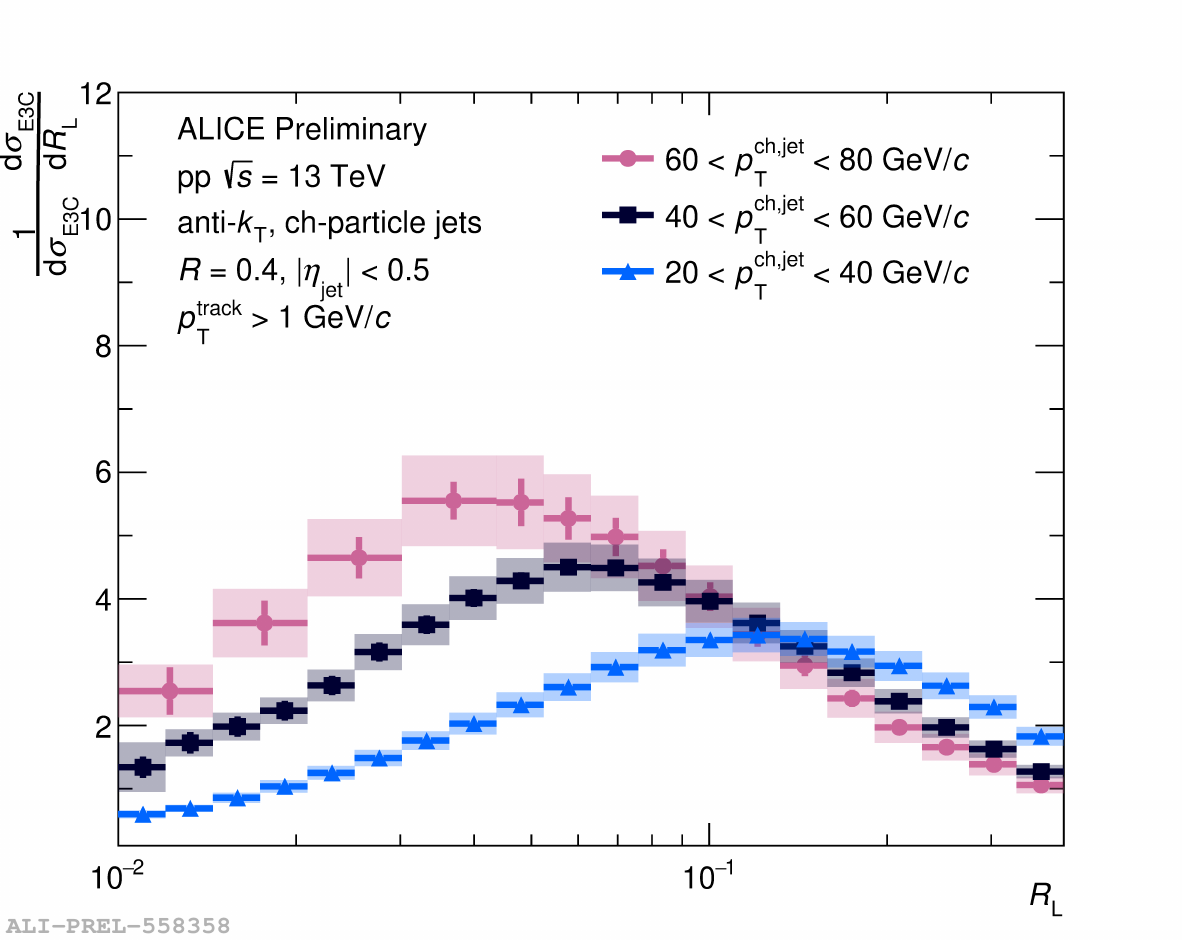}
    \end{subfigure}
    \begin{subfigure}{0.49\textwidth}
        \centering
        \includegraphics[scale=.35]{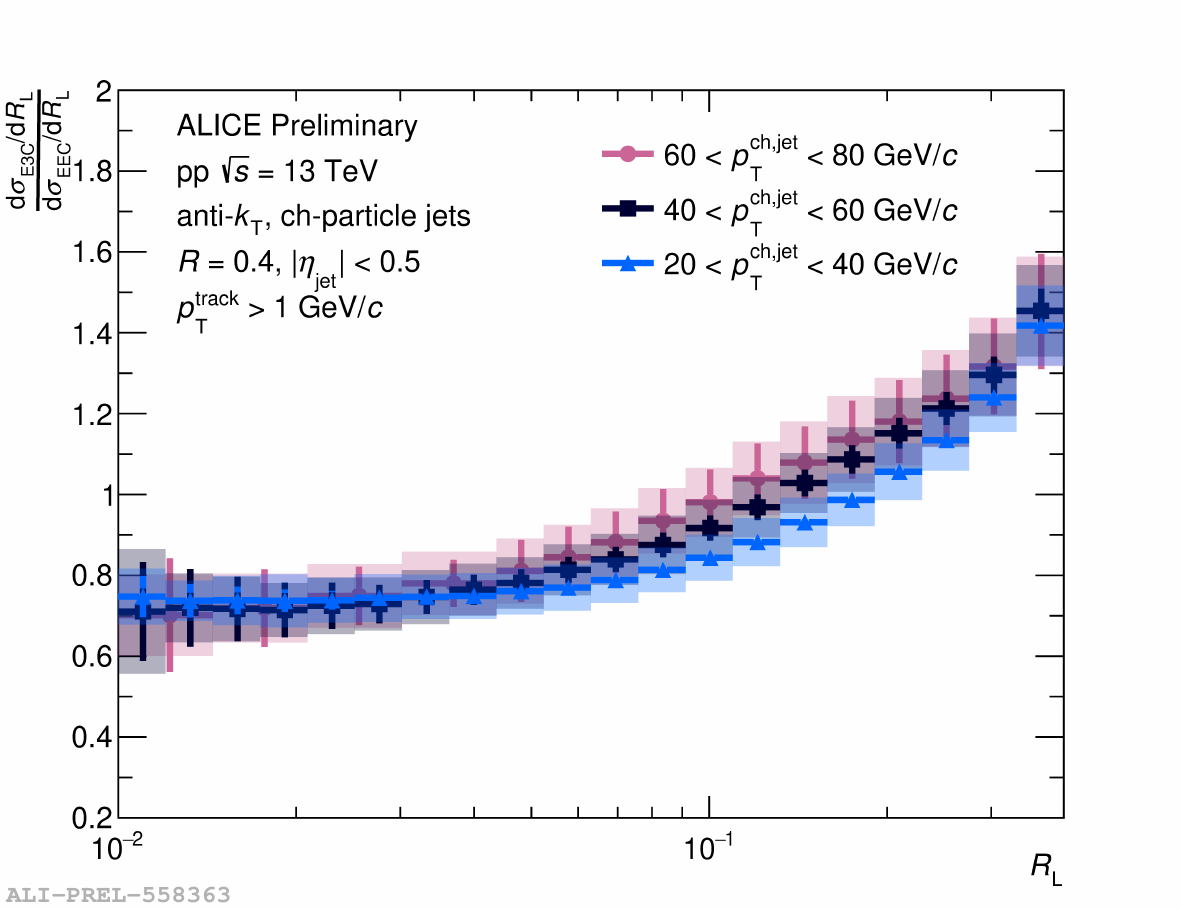}
    \end{subfigure}
    \caption{\textbf{(a)} The E3C for inclusive jets in pp collisions in three different jet \pt~ranges. \textbf{(b)} The ratio of E3C/EEC, shown in the same jet \pt~ranges.} 
    \label{fig:e3c_pp}
\end{figure}

ALICE measured the E3C in pp collisions, shown in Fig.~\ref{fig:e3c_pp}(a). While the E3C appears to have very similar qualitative features to the EEC, the slopes in the large \rl~region depend on different anomalous dimensions. The 2-point EEC depends on the anomalous dimension of the local twist-2 spin-3 operator, while the E3C depends on the twist-2 spin-4 operator. By taking the ratio of the E3C/EEC, the anomalous dimensions can be used to extract $\alpha_s$. \cite{e3c_theory}

The ratio of the E3C to EEC can be seen in Fig.~\ref{fig:e3c_pp}(b). The slope in the ratio at wide angles is expected to be proportional to $\alpha_s \ln(R_L)$ and changes for different jet energies, indicating sensitivity to the running of the strong coupling constant. ALICE is currently working to extract the value of $\alpha_s$ in complementary phase space to the recent CMS measurement. \cite{cms_e3c}

\section{Conclusion}

In summary, the energy correlators are a novel tool to study jet formation and evolution, and show a separation between the perturbative and non-perturbative regions. ALICE has performed a number of EEC measurements in various collision systems and types of jets. The EECs are sensitive to both mass effects and flavor effects, as shown by the \Dzero-tagged jet measurement. When studying jets in p--Pb collisions, an interesting modification at low jet energies is apparent, and is being studied further. Finally, the E3C currently provides a precise way of extracting $\alpha_s$ using jet substructure techniques. Looking forward, a number of additional energy correlator measurements are in progress and being analyzed to further constrain the underlying physics of jet evolution, hadronization, and the quark-gluon plasma.

\section*{References}
\bibliography{moriond}

\begin{thebibliography}{10}

\bibitem{eec_intro_chen}
H.~Chen, I.~Moult, X.~Zhang, and H.~X. Zhu.
\newblock {\em Phys. Rev. D}, 102:054012, Sep 2020.

\bibitem{alice_eec_pp}
ALICE Collaboration.
\newblock [arXiv:2409.12687 [hep-ex]].

\bibitem{alice_deadcone}
ALICE Collaboration.
\newblock {\em Nature}, 605:440--446, 2022.

\bibitem{alice_eec_djets}
ALICE Collaboration.
\newblock [arXiv:2504.03431 [hep-ex]].

\bibitem{eec_croft_pythia}
E.~Craft, K.~Lee, B.~Meçaj, and I.~Moult.
\newblock [arXiv:2210.09311 [hep-ph]].

\bibitem{cms_rho_sub}
CMS Collaboration.
\newblock {\em J. High Energ. Phys.}, 2012:130.

\bibitem{ppb_theory_fu}
Y.~Fu, B.~Müller, and C.~Sirimanna.
\newblock [arXiv:2411.04866 [nucl-th]].

\bibitem{ppb_theory_barata}
J.~Barata, Z.-B. Kang, X.~M. López, and J.~Penttala.
\newblock [arXiv:2411.11782 [hep-ph]].

\bibitem{ppb_theory_andres}
C.~Andres, F.~Dominguez, J.~Holguin, C.~Marquet, and I.~Moult.
\newblock [arXiv:2411.15298 [hep-ph]].

\bibitem{anjali_QM25_slides}
A.~Nambrath.
\newblock [https://indico.cern.ch/event/1334113/contributions/6350975/].

\bibitem{e3c_theory}
H.~Chen, M.-X. Luo, I.~Moult, T.-Z. Yang, X.~Zhang, and H.~X. Zhu.
\newblock {\em Journal of High Energy Physics}, 2020(8), Aug 2020.

\bibitem{cms_e3c}
CMS Collaboration.
\newblock {\em \PRL}, 133:071903, Aug 2024.

\end{thebibliography}

\end{document}